\documentclass[acmlarge]{acmart}
\setcopyright{none}
\acmDOI{}
\copyrightyear{} 
\acmYear{}
\AtBeginDocument{%
  }

\usepackage[ruled,vlined]{algorithm2e}
\DeclareMathOperator*{\argmin}{argmin} 
\usepackage{amsmath} 
\usepackage{multirow}
\usepackage{booktabs}
\usepackage{tabularx}  

\begin{document}

\title{TRiMM: Transformer-Based Rich Motion Matching for Real-Time multi-modal Interaction in Digital Humans}

\author{Yueqian Guo}

\email{xiaoheguo@outlook.com}
\affiliation{%
  \institution{Jiangxi University of Finance and Economics}
  \country{China}
  \city{Nanchang}
}

\author{Tianzhao Li}

\email{CUCLelivre@gmail.com}
\affiliation{%
  \institution{Communication University of China}
  \country{China}
  \city{Beijing}
}

\author{Xin Lyu}

\email{lvxinlx@cuc.edu.cn}
\affiliation{%
  \institution{Communication University of China}
  \country{China}
  \city{Beijing}
}

\author{Jiehaolin Chen}

\email{1202390090@jxufe.edu.cn}
\affiliation{%
  \institution{Communication University of China}
  \country{China}
  \city{Beijing}
}

\author{Zhaohan Wang}

\email{1202390090@jxufe.edu.cn}
\affiliation{%
  \institution{Communication University of China}
  \country{China}
  \city{Beijing}
}

\author{Sirui Xiao}

\email{nightcruising79@gmail.com}
\affiliation{%
  \institution{Communication University of China}
  \country{China}
  \city{Beijing}
}

\author{Yurun Chen}

\email{2575738708@qq.com}
\affiliation{%
  \institution{Communication University of China}
  \country{China}
  \city{Beijing}
}

\author{Yezi He}

\email{yates00619@gmail.com}
\affiliation{%
  \institution{Communication University of China}
  \country{China}
  \city{Beijing}
}

\author{Helin Li}

\email{kingselyee67@sina.cn}
\affiliation{%
  \institution{Communication University of China}
  \country{China}
  \city{Beijing}
}

\author{Fan Zhang}
\authornote{corresponding author}
\email{Fanzhang@cuz.edu.cn}
\affiliation{%
  \institution{Communication University of Zhejiang}
  \country{China}
  \city{Beijing}
}


\begin{abstract}
Large Language Model (LLM)-driven digital humans have sparked a series of recent studies on co-speech gesture generation systems. However, existing approaches struggle with real-time synthesis and long-text comprehension. This paper introduces Transformer-Based Rich Motion Matching (TRiMM), a novel multi-modal framework for real-time 3D gesture generation. Our method incorporates three modules: 1) a cross-modal attention mechanism to achieve precise temporal alignment between speech and gestures; 2) a long-context autoregressive model with a sliding window mechanism for effective sequence modeling; 3) a large-scale gesture matching system that constructs an atomic action library and enables real-time retrieval. Additionally, we develop a lightweight pipeline implemented in the Unreal Engine for experimentation. Our approach achieves real-time inference at 120 fps and maintains a per-sentence latency of 0.15 seconds on consumer-grade GPUs (Geforce RTX3060). Extensive subjective and objective evaluations on the ZEGGS, and BEAT datasets demonstrate that our model outperforms current state-of-the-art methods. TRiMM enhances the speed of co-speech gesture generation while ensuring gesture quality, enabling LLM-driven digital humans to respond to speech in real time and synthesize corresponding gestures. Our code is available at \url{https://github.com/teroon/TRiMM-Transformer-Based-Rich-Motion-Matching}
\end{abstract}

\keywords{Transformer-Based Models, Multi-modal Fusion, Large motion graph, Real-Time Motion Generation, Digital Humans}


\maketitle

\section{Introduction}
\begin{figure}[H]
  \centering
  \includegraphics[width=0.4\linewidth]{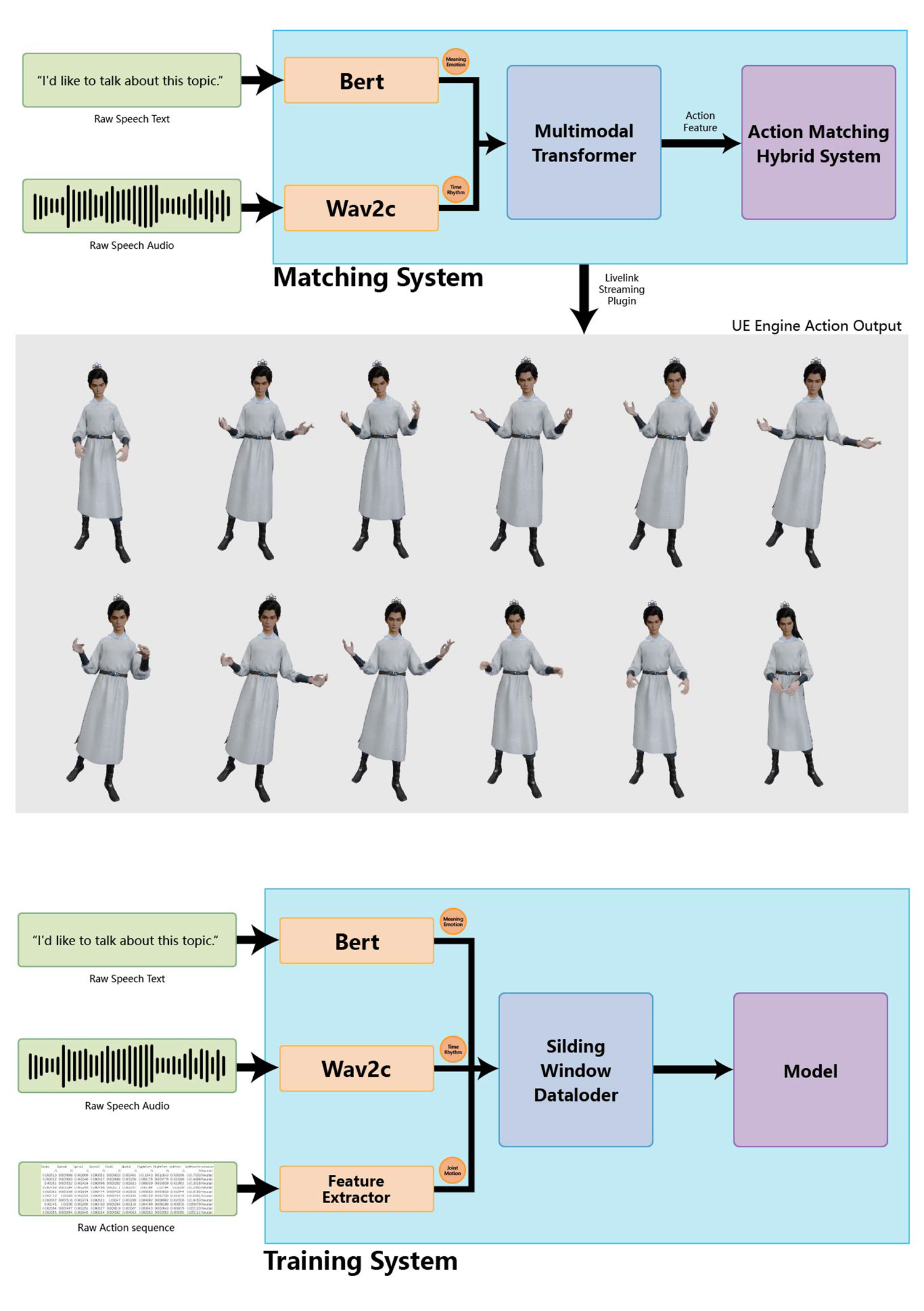}
  \caption{The system trains a Transformer model using text, speech, and motion feature data, performs real-time multi-modal inference with text and speech to predict motion features, synthesizes actions in real-time from a large motion library via Motion Matching, and renders the results through Unreal Engine using LiveLink.}
  \Description{System Overview}
\end{figure}

Recent advances in Human-computer interaction have demonstrated the potential of large language models (LLMs) to drive digital human behaviors\cite{wan_building_2024}\cite{sonlu_effects_2024}\cite{wu2025motionagent}. Therefore, many methods for generating digital human actions have emerged recently. While current methods excel at text-based dialogue generation, the integration of real-time motion synthesis remains a critical challenge for immersive applications like virtual live-streaming and interactive gaming\cite{chen_choreomaster_2021}\cite{tang_real-time_2022}\cite{xiao_motionstreamer_2025}
Traditional approaches for digital human animation rely on two distinct paradigms: motion capture-based systems and deep generative models. Early motion matching techniques \cite{clavet2016motion} leveraged pre-recorded motion libraries to ensure natural movements, but suffered from limited action diversity due to combinatorial constraints. Recent generative approaches using VAEs\cite{petrovich_action-conditioned_2021}\cite{yao_controlvae_2022}\cite{petrovich_temos_2022}and diffusion models \cite{zhang2024motiondiffuse}\cite{zhang_remodiffuse_2023}\cite{karunratanakul_guided_2023} have improved motion variety, yet their prohibitive computational demands hinder real-time deployment  MDM \cite{tevet_human_2022}requires 12s to produce a motion sequence given a textual description, making it impractical for interactive applications.

This paper introduces Transformer-Based Rich Motion Matching (TRiMM) that addresses three fundamental limitations in existing methods: (1) Temporal misalignment between speech prosody and gesture dynamics in cross-modal synthesis; (2) Restricted motion diversity caused by limited action libraries and rigid retrieval mechanisms; (3) Computational overhead from iterative denoising processes in diffusion-based approaches.

Our key innovations include:
\begin{itemize}
\item A multi-modal attention mechanism that dynamically aligns speech and text features with gesture kinematics. it is implemented through space-time attention transformer, achieving precise synchronization between speech prosody and gesture dynamics.

\item A sliding-window autoregressive model that maintains 8 sentences' contextual memory. it also achieves 0.159s inference latency, addressing the computational overhead problem in diffusion-based approaches. Our method enables real-time motion synthesis at 120 fps on RTX3060.

\item A hierarchical motion graph containing 9,143 atomic actions with multi-criteria similarity search (semantic, kinematic, temporal). it overcomes the limited action diversity in traditional motion capture systems,and raises motion diversity by 2 times.
\end{itemize}

\section{Related Work}
\subsection{Generative Methods}

The main methods, contributions, and limitations of the current diffusion model in the field of action generation are as follows: motiondiffusion, as a text-driven action generation framework based on the diffusion model, realizes diversified fine-grained action synthesis through probabilistic mapping and hierarchical operation. Its core contribution is to introduce the de-noising process instead of deterministic mapping, which breaks through the limitation of action diversity of traditional retrieval methods. Experiments show that its generation quality exceeds the same period method; However, its multi-step iterative generation mechanism leads to significant reasoning delay, which can not meet the needs of real-time interaction. MDM \cite{tevet_human_2022} constructs motion latent space through a two-stage training strategy, and has made progress in improving the generation efficiency. However, the non-end-to-end training mode leads to incomplete motion distribution modeling, and it is difficult to accurately capture complex motion patterns. MLD \cite{chen_executing_2023}uses the latent space diffusion strategy to reduce the computational complex motion sequence compression through hierarchical feature extraction. However, its two-stage training architecture has the problem of insufficient feature decoupling, resulting in the loss of motion details.  \cite{zhang_remodiffuse_2023}innovatively integrates the retrieval strategy and diffusion model to enhance the semantic consistency of generated actions through semantic similarity matching. However, the introduction of retrieval mechanism increases the computational overhead and fails to fundamentally solve the real-time bottleneck. \cite{leonardis_emdm_2025} accelerates the generation process by improving the sampling algorithm, but its optimization direction focuses on the algorithm level, and lacks the explicit modeling of physical constraints (such as foot contact dynamics), which may lead to the generation action violating the biomechanical laws.

It is worth noting that although transformer architecture shows potential in time series modeling, the existing methods have not fully exploited the advantages of the autoregressive mechanism, which enhances the model's comprehension of long-text dialogues in LLM-driven conversations, enabling coherent gesture generation across extended interactions. Most generative models suffer from high computational complexity, making it challenging to achieve real-time inference above 30 fps. Our TRiMM framework addresses this limitation by combining sliding window autoregressive generation with a lightweight deployment pipeline, enabling real-time motion synthesis at 120 fps with a per-sentence latency of only 0.13 seconds on consumer-grade GPUs.

\subsection{Retrieval-based Methods}

The retrieval-based motion generation method continues to evolve in the field of animation and gesture synthesis. Motion Matching \cite{clavet2016motion}  aims to retrieve the optimal next frame animation from the pre-recorded animation library. This technology replaces the manual state setting of traditional state machines in the motion capture database animation. However, it has two major limitations: memory usage increases linearly with the amount of animation data, and the fixed feature weights lead to the need for manual adjustment of the matching strategy in complex interaction scenarios. \cite{holden_learned_2020} divides the traditional motion-match algorithm into three stages: projection, stepping, and decompression, which are replaced by neural networks, respectively, significantly reducing the memory dependence and improving the scalability. However, the physical rationality of the generated action still depends on the quality of the original action in the database, which is difficult to deal with the no motion mode.\cite{habibie_motion_2022} introduces audio pose joint similarity retrieval in speech-driven gesture synthesis, and optimizes the retrieval results in combination with confrontation generation network. Its contribution is to improve the synchronization and naturalness of motion and voice rhythm; However, this method is still based on keyword-level semantic matching and lacks a deep understanding of context semantics, resulting in a rough association between gesture and semantics in complex conversation scenes. \cite{yang_qpgesture_2023} quantized the discrete gesture units through the gesture vq-vae module, aligned the speech and gesture sequences using Levenshtein Distance, and introduced phase guidance to optimize the motion matching, which effectively alleviated the problem of random motion jitter and phoneme asynchrony; However, its two-stage quantization matching process introduces additional computational delay, and relies on a fixed size of pre training codebook, which is difficult to dynamically expand diversified actions. Although the current retrieval methods have advantages in real-time and physical rationality, they are still limited by a single-level index structure (such as pure motion features or phoneme matching), resulting in coarse semantic granularity and insufficient accuracy of cross-modal alignment. 

In this regard, our proposed hierarchical action retrieval engine uses a semantic action phoneme three-level index architecture to simultaneously model language context, action coherence and phoneme timing constraints in real-time retrieval, support the dynamic combination of 9143 atomic actions, and break through the bottleneck of the capacity and alignment accuracy of the traditional retrieval library.
\subsection{multi-modal Fusion Approaches}

In the latest progress in the field of multi-modal driven action generation,\cite{DBLP:journals/corr/abs-2403-02905} fused multi-modal signals such as voice, emotion, and identity through a diffusion model, proposed a progressive intermediate fusion strategy and a mask style matrix to achieve fine-grained voice action alignment. Its core contribution is to use the geometric loss function to constrain the continuity of joint velocity and acceleration to generate a high smooth motion sequence; However, its multi-modal interaction relies on the static intermediate fusion mechanism, and the dynamic context (such as the semantic evolution in the dialogue History) is not fully modeled, resulting in the rigid switching of action style in the long-term generation. \cite{zhang_large_2024} Integrates the body part perception modeling into the diffusion transformer backbone network through the unified dataset motionverse and artattention attention mechanism, supports the joint control of multi-modal inputs such as text and music, and breaks through the generalization limitation of single task model; However, its pre training strategy relies on fixed frame rate and mask mode, which is difficult to adapt to the asynchronous timing changes of multi-modal signals in real-time interactive scenes. Based on the coarse-to-fine training strategy and MC Attn parallel human body topology modeling, \cite{bian2025motioncraft} realizes the zoning control of whole body movements (such as the decoupling of fine hand movements and trunk posture). Its contribution is to enhance the generation controllability through text semantic pre-training and low-level control adaptation; However, its phased training framework leads to the response delay of dynamic control signals (such as real-time user instructions), and multi-position independent control may destroy the physical consistency of the kinematic chain. Although the existing methods have made progress in the breadth of multi-modal fusion, the temporal modeling of dynamic context (such as real-time bullet screen semantics in virtual live broadcast) is still limited to window-level splicing, lacking the dynamic weight allocation mechanism of cross-modal features. 

In this regard, we propose a multi-modal context fusion transformer, which dynamically aligns language semantics (such as real-time dialog text) and motion features through cross modal attention, and realizes the joint reasoning of voice rhythm, text emotion and user intention at the transformer level, effectively solving the problems of semantic gesture dislocation and style discontinuity in the generation of virtual humanoid actions.

\section{Method}
The Transformer-Based Rich Motion Matching (TRiMM) system is composed of five core modules. 1) The multi-modal feature extractor leverages WAVE2vec2 and Bert to extract audio and text features. 2) The feature encoder-decoder fuses these features via a gated mechanism and enables autoregressive prediction with a sliding window. 3) The Time-Space attention transformer captures space-time relationships using positional encoding and attention mechanisms. 4) The Real-Time Motion Matching Engine constructs a K-NN graph for action retrieval. 5)  the motion hybrid system uses quaternion-based interpolation for smooth motion transitions. Together, these components enable TRiMM to process multi-modal data effectively and generate accurate motion predictions. 

\begin{figure}[H]
  \centering
  \includegraphics[width=0.5\linewidth]{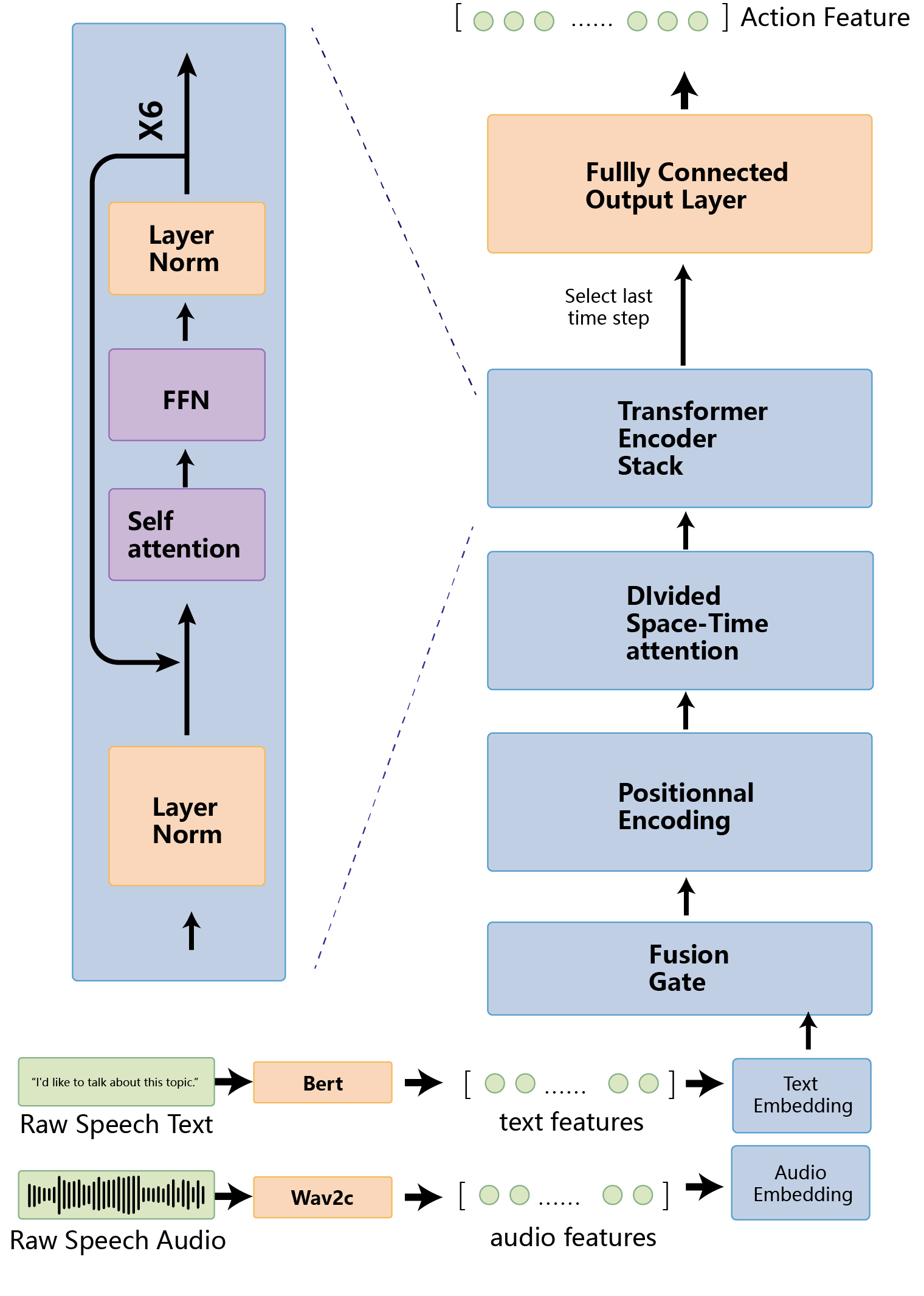}
  \caption{This system processes text and audio inputs through dedicated embedding layers, combines them via a fusion gate, and adds positional encoding to capture timing. The fused features then pass through a divided space-time attention mechanism and a 6-layer Transformer encoder stack (with layer normalization, self-attention, and feed-forward networks) to model space-time relationships. The final time step is selected and mapped by a fully connected output layer to predict the next-frame action feature.}
  \Description{The Architecture of TRiMM}
\end{figure}
\subsection{Multi-modal information feature extractor}

In view of the multi-modal characteristics and autoregressive properties\cite{katharopoulos_transformers_2020} of this study, this study uses \cite{schneider_2019_wav2vec} and \cite{devlin2019bert} models to extract the feature vectors of audio and text respectively, and uses the sliding window mechanism to input serialized data to the transformer model, so as to learn the context information existing in the text and audio sequence.

In this study, WAVE2 vec2 base 960h model is used to extract audio features. It can extract a two-dimensional matrix with timing spectrum characteristics from the original audio. To unify the feature dimensions, we use principal component analysis (PCA)\cite{greenacre2022principal} to reduce the dimension of the feature matrix to a fixed 2048 dimensional vector to form a standardized audio feature representation. Because wav2vec can learn the context information between different time steps in audio, it can be combined with the powerful long sequence modeling ability and multi head attention mechanism of transform model to significantly improve the speech comprehension ability of the model.

Text feature extraction is based on Bert model. By extracting the final hidden state (768 dimensions) of the special tag [cls] as the text representation, the vector contains the deep semantic information of the sentence.

\subsection{Encoder and Decoder of feature vector}
\subsubsection{Feature encoder and multi-modal Embedding Fusion}

The input text features $\mathbf{T} \in \mathbb{R}^{d_t}$ ($d_t=768$) and audio features $\mathbf{A} \in \mathbb{R}^{d_a}$ ($d_a=512$) undergo a three-stage gated fusion process:

Unified Representation Learning:  
Both modalities are first aligned into a shared metric space using dedicated projection layers. The text features are transformed to a 1024-dimensional subspace through a learnable linear layer, while the audio features undergo similar dimensionality expansion. This resolves distribution discrepancies between modalities.

Context-Aware Gating:  
A gating mechanism dynamically adjusts modality contributions by learning adaptive weights. The concatenated projected features pass through a sigmoid-activated layer that generates two gating coefficients ($g_t$, $g_a$), reflecting the relative importance of each modality for the current input.

Adaptive Fusion:  
The final representation combines the projected features through gated summation:  
$\mathbf{H}_{\text{fusion}} = g_t \cdot \mathbf{T}' + g_a \cdot \mathbf{A}' \in \mathbb{R}^{1024}$  
followed by duplication to form the 2048D output $\mathbf{H}_{\text{final}}$. This architecture enables automatic calibration of modality weights while preserving information density.

The function of embedding and fusion gate mechanism is to fuse text and speech in two modalities, and the difference in distribution between modalities can be solved by linear projection of embedding layers into a unified metric space. Through the gating mechanism, the modal weights in different scenarios can be automatically adjusted.

\subsubsection{Action decoder and autoregressive prediction }
In this study, a sliding window mechanism was used to read multiple previous text and speech feature frames. Predicts the action frame at the next time, enabling autoregressive prediction. This mechanism enhances the model's contextual understanding of text and speech. 

\begin{figure}[H]
  \centering
  \includegraphics[width=0.5\linewidth]{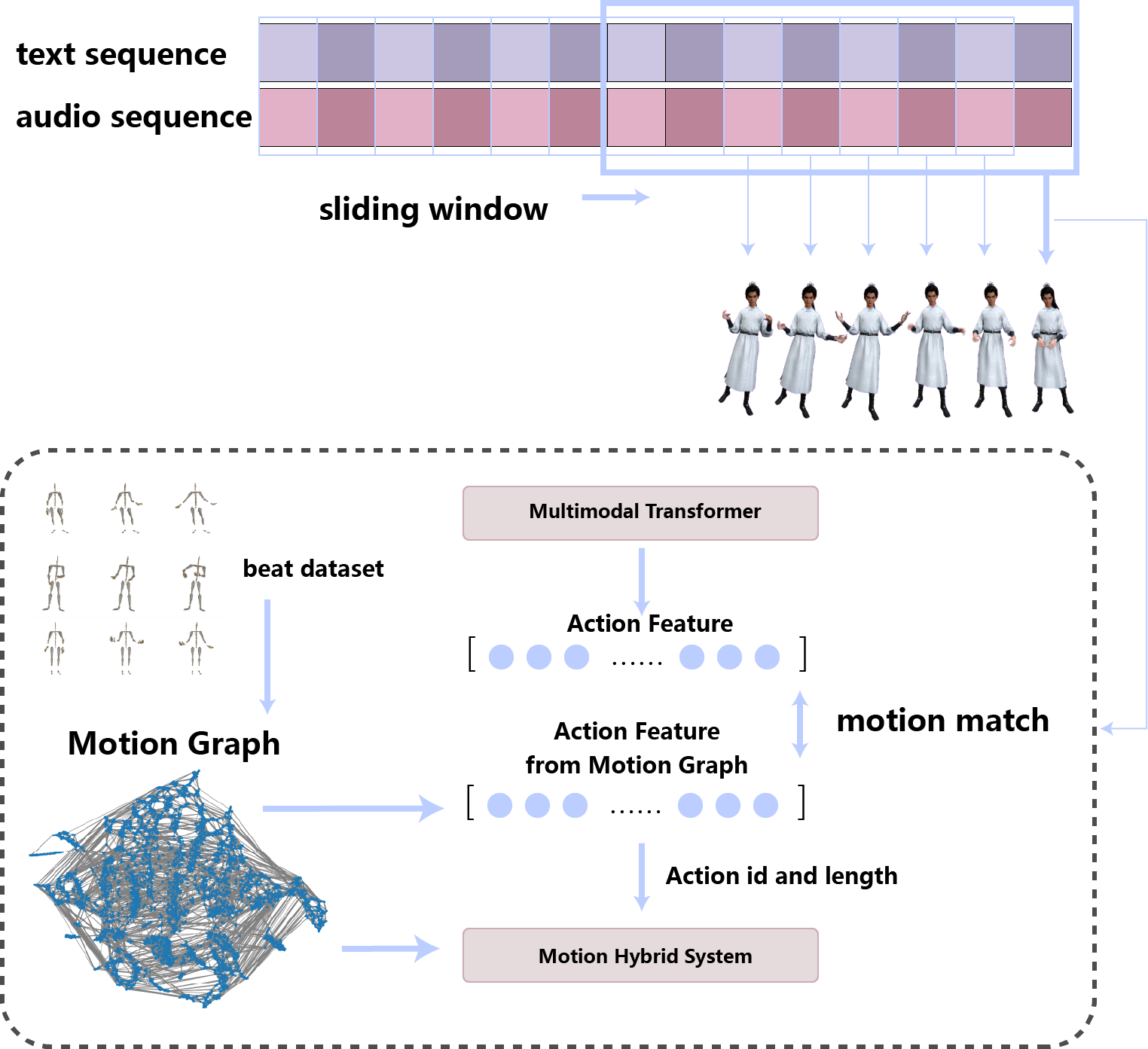}
  \caption{This figure illustrates the autoregressive prediction process. And the connection between autoregressive prediction, motion match engine, and motion hybrid system.}
  \Description{The Architecture of autoregressive prediction}
\end{figure}

We use the \cite{lee_2001_sliding} strategy to process the text and audio feature sequences separately and generate sequential feature vector information according to the first several frames of the current time point. By inputting the feature vector in the sliding window into the transformer, the model can learn the time or order relationship between features, so that it can predict the next element in the next window or sequence according to the features in the current window and the information of the previous window, so as to realize autoregressive prediction.
Let $N$ be the window size. At timestep $t$, the input matrix contains $N$ historical multi-modal features:

\begin{equation}
\mathbf{X}_t = \begin{bmatrix} 
\mathbf{T}_{t-N+1} & \mathbf{A}_{t-N+1} \\
\vdots & \vdots \\
\mathbf{T}_t & \mathbf{A}_t 
\end{bmatrix} \in \mathbb{R}^{N \times (d_t + d_a)}
\end{equation}

where $\mathbf{T}_k \in \mathbb{R}^{d_t}$ and $\mathbf{A}_k \in \mathbb{R}^{d_a}$ represent text and audio features at timestep $k$.

In the autoregressive process, the transformer uses the previous prediction results as the conditions for subsequent prediction, which can make the generated output more coherent and logical. Moreover, the sliding window can divide the feature sequences with different lengths into fixed length windows, which gives the transformer model the ability to adapt to different length sequences.
The encoder of the action feature selects the last time step of the Transformer Encoder stack and outputs the 750-dimensional action feature value through a fully connected layer.
\begin{equation}
\begin{split}
  \text{Transformer Output:} & \quad 
  \mathbf{C}^* = \mathrm{TransformerBlock}^L(\mathbf{C}) \\
  \text{Action Prediction:} & \quad 
  \hat{\mathbf{Y}}_{t+1} = \mathbf{W}_o\mathbf{C}^*[-1,:] \\
  \text{Window Update:} & \quad 
  \mathbf{X}_{t+1} = \begin{bmatrix}
  \mathbf{X}_t[1:,:] \\
  (\mathbf{T}_{t+1}, \mathbf{A}_{t+1})
  \end{bmatrix}
\end{split}
\end{equation}

\subsection{Time-Space attention transformer}
\subsubsection{Positional Encoding }
To preserve the temporal sequence, we augment each patch vector with positional embeddings, which encode the spatial and temporal location of the patch within the audio-text sequence. Positional embeddings follow standard sinusoidal encoding.
The encoded features become:

\begin{equation}
\tilde{X} = X + \text{PE}(1:T) \in \mathbb{R}^{T \times d_e}
\end{equation}

This allows the model to learn relative positions via trigonometric identities:

\begin{equation}
\text{PE}(pos+\Delta) = \text{PE}(pos)\cdot\text{PE}(\Delta) + \text{cross-terms}
\end{equation}

\subsubsection{Temporal Spacial Attention Mechanism}

In this study, we employ Space-Time Attention within a Transformer-based architecture to facilitate action inference from integrated textual and vocal data. This approach allows our model to capture the intricate relationships between multi-modal inputs across both spatial and temporal dimensions, enabling the prediction of subsequent action states.

The input sequence first calculates self attention on the time dimension:

\begin{equation}
\begin{aligned}
& X_{\text{time}} = X \in \mathbb{R}^{B \times S \times D} \\
& \{Q,K,V\}_{\text{time}} = \text{Permute}(X_{\text{time}}) \in \mathbb{R}^{S \times B \times D} \\
& \text{Output}_{\text{time}} = X + \text{Permute}^{-1}(\text{Attention}(Q,K,V))
\end{aligned}
\end{equation}

Where attention weights are computed via scaled dot-product similarity.

After feature transformation, attention is paid on the transformed spatial feature dimension:

\begin{equation}
  \begin{aligned}
& X_{\text{space}} = W_{\text{proj}}X \in \mathbb{R}^{B \times S \times D} \\
& \{Q,K,V\}_{\text{space}} = \text{Permute}(X_{\text{space}}) \\
& \text{Output}_{\text{final}} = \text{LayerNorm}(X_{\text{space}} + \text{Permute}^{-1}(\text{Attention}(Q,K,V)))
  \end{aligned}
\end{equation}

The model computes attention weights for each patch by measuring the dot-product similarity between query, key, and value representations. This process allows the model to selectively focus on the most relevant patches when predicting the next state in the action sequence.

\subsection{Real-Time Motion Matching Engine}
\subsubsection{K-NN Graph Construction  }

The construction process of the action graph can be mathematically formalized as follows: Given a set of \( n \) actions, each action \( a_i \) (\( i \in \{1, 2, \ldots, n\} \)) is characterized by two attributes: feature vector \( x_i \in \mathbb{R}^d \) (\( d = 750 \)) and duration scalar \( t_i \in \mathbb{R}^+ \). We define a metric space \( (\mathbb{R}^d, d) \) using the Euclidean distance. For each \( x_i \), compute its \( k \)-nearest neighbors (k-NN) set:

\begin{equation}
\mathcal{N}_k(x_i) = \left\{x_j \in X \, \middle| \, \text{\( x_j \) is one of the \( k \) closest points to \( x_i \) under metric \( d \)} \right\}.
\end{equation}

Construct an undirected weighted graph \( G = (V, E, W) \):  
Vertex set: \( V = \{v_1, v_2, \ldots, v_n\} \), where \( v_i \leftrightarrow a_i \).  

Edge set:  

\begin{equation}
E = \bigcup_{i=1}^n \left\{(v_i, v_j) \, \middle| \, x_j \in \mathcal{N}_k(x_i) \land i \neq j \right\}.
\end{equation}

Edge weight function:  

\begin{equation}
w_{ij} = \|x_i - x_j\|_2 = \sqrt{\sum_{k=1}^d (x_{i,k} - x_{j,k})^2}, \quad \forall (v_i, v_j) \in E.
\end{equation}

\begin{figure}[H]
  \centering
  \includegraphics[width=0.4\linewidth]{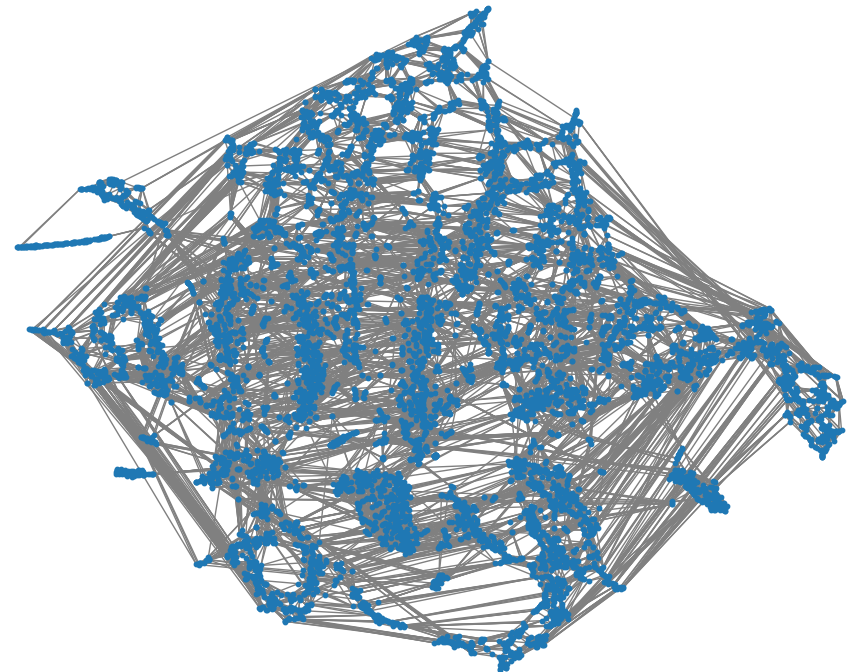}
  \caption{K-NN Graph Visualization after t-SNE Dimensionality Reduction. The graph consists of 9,143 action nodes, where each node contains an action feature vector and its corresponding duration, with every node connected to its K=10 nearest neighboring nodes to form a K-Nearest Neighbor (K-NN) graph structure.}
  \Description{K-NN Graph Visualization after t-SNE Dimensionality Reduction}
\end{figure}
This graph structure establishes a topological foundation for efficient graph-traversal-based action sequence retrieval and optimization.

\subsubsection{Retrieval Pipeline}

The algorithm begins by identifying the nearest node in the preconstructed KNN graph \cite{zhao_approximate_2020}to the previous action feature vector, which serves as the starting point for constrained search. 

It then retrieves the top-k neighbors of this node, sorted by their Euclidean distance to the previous vector, to prioritize locally similar actions. A breadth-first search (BFS)\cite{kurant_bias_2010} is initiated from these neighbors, expanding iteratively to their connected nodes while avoiding revisiting explored nodes. During traversal, the algorithm enforces a duration constraint by only considering nodes whose precomputed action durations exceed a threshold t For each valid node, the Euclidean distance between its feature vector and the current target vector is calculated and stored. 

The search terminates once the queue is exhausted, and the node with the smallest distance to the current vector is selected as the optimal match. If no nodes satisfy the duration constraint during the entire traversal, the algorithm returns an empty result. This approach balances motion continuity (via graph-based proximity to the previous action) and temporal constraints while efficiently exploring the graph structure to align with evolving action requirements.
\begin{algorithm}
  \caption{Constrained KNN Search with Duration Filter}
  \KwIn{prev vector, current vector, graph $G$, vectors $V$, action durations $D$, action duration $\tau$, top k}
  \KwOut{best action index or $\emptyset$}

  $prev nearest idx \leftarrow \argmin_{i} \|V[i] - prev vector\|_2$\;
  
  $prev neighbors \leftarrow$ Sort $G[prev nearest idx]$ by $\|V[x] - prev vector\|_2$\;
  $prev neighbors \leftarrow prev neighbors[1..top k]$\;
  
  Initialize visited set: $\mathcal{V} \leftarrow \emptyset$\;
  Initialize queue $\mathcal{Q}$ with $prev neighbors$\;
  Initialize valid nodes: $\mathcal{N} \leftarrow [\,]$, $\mathcal{D} \leftarrow [\,]$\;
  
  \While{$\mathcal{Q} \neq \emptyset$}{
    $u \leftarrow \mathcal{Q}$.pop(0)\;
    \If{$u \in \mathcal{V}$}{
      continue\;
    }
    $\mathcal{V} \leftarrow \mathcal{V} \cup \{u\}$\;
    
    \If{$D[u] > \tau$}{
      $d \leftarrow \|V[u] - current vector\|_2$\;
      $\mathcal{N}$.append(u), $\mathcal{D}$.append(d)\;
    }
    
    \tcp{Add neighbors of $u$ to queue}
    $\mathcal{Q} \leftarrow \mathcal{Q} \cup G[u]$\;
  }
  
  \If{$\mathcal{N} \neq \emptyset$}{
    $best index \leftarrow \argmin_{d} \mathcal{D}$\;
    \Return $\mathcal{N}[best index]$\;
  }
  \Else{
    \Return $\emptyset$\;
  }
\end{algorithm}

\subsection{Motion Hybrid system}

We propose a real-time gradual change switching method of motion signal based on quaternion, We realize the seamless transition of BVH animation sequence through the dual mechanism of quaternion rotation representation and cube interpolation.
In the attitude interpolation phase, the quaternion spherical linear interpolation algorithm is adopted:
\begin{equation}
\mathbf{q}(t) = \frac{\sin((1-t)\theta)}{\sin\theta}\mathbf{q}_1 + \frac{\sin(t\theta)}{\sin\theta}\mathbf{q}_2
\end{equation}
Where $\theta$ is the angle between quaternion $\mathbf{q}_1$ and $\mathbf{q}_2$, $t$  is the interpolation parameter. The formula ensures the shortest path interpolation of the rotation axis on the spherical surface, and effectively avoids the problem of universal joint deadlock caused by Euler angle interpolation.
The position gradient uses cube interpolation function to realize smooth transition of continuous acceleration:
\begin{equation}
L(t) = (2t^3-3t^2+1)L_1 + (-2t^3+3t^2)L_2
\end{equation}

Where $l_1$and $l_2$respectively represent the position coordinates of adjacent keyframes. Its speed curve meets:

\begin{equation}
v(t) = \frac{dL}{dt} = (6t^2-6t)L_1 + (-6t^2+6t)L_2
\end{equation}
The interpolation function satisfies the zero boundary condition (V (0)=v (1)=0) at the endpoint to ensure continuous speed at the beginning and end of the switching action, which effectively inhibits the sudden change of mechanical motion.

\subsection{Training and inference}
The training process involves the following steps:
Formally, for a window size W and input sequences:

Text: $X_{\text{text}} = [x_t, x_{t+1}, \ldots, x_{t+W-1}]$ , extracted from textual data using Bert.

Audio: $X_{\text{audio}} = [a_t, a_{t+1}, \ldots, a_{t+W-1}]$ , extracted from audio data using WAVE2vec2.

True Action: $y = [y_t, y_{t+1}, \ldots, y_{t+W-1}]$, extracted from Bvh files. We convert Bvh files into numpy arrays and then use PCA to reduce the dimensionality.

The model predicts: $\hat{y} = f(X_{\text{text}}, X_{\text{audio}}) \to y_{t+W-1}$ , representing the predicted action feature. 

For loss function, we use Mean Squared Error (MSE) loss between predicted and true actions:
\begin{equation}
L(\theta) = \frac{1}{N} \sum_{i=1}^N \|\hat{y}_i - y_i\|_2^2
\end{equation}
where $\theta$ represents all trainable parameters

For optimizer, we use Adam optimizer to update parameters via:
\begin{equation}
\theta_t = \theta_{t-1} - \eta_t \hat{m}_t/(\sqrt{\hat{v}_t} + \epsilon)
\end{equation}
where $\hat{m}_t$ and $\hat{v}_t$ are bias-corrected momentum estimates,
$\eta_t$ is the decaying learning rate

The sliding window creates the training pairs \((X_{\text{text}}^{t:t+W}, X_{\text{audio}}^{t:t+W}) \to y^{t+W}\), allowing the model to learn temporal dynamics while maintaining fixed-length input dimensionality. The attention mechanism then discovers relevant cross-modal patterns within these windows, while the optimization process adjusts parameters to minimize prediction error across the dataset.

In inference processing, audio and text inputs are processed through a multi-modal transformer to generate motion features, which are then synthesized into real-time animations via a motion matching system. These animations are converted into JSON signals and streamed through the LiveLink plugin to be rendered with MetaHuman characters in Unreal Engine 5. In the mean time motion signal is recorded in the format BVH.
\begin{figure}[H]
  \centering
  \includegraphics[width=0.5\linewidth]{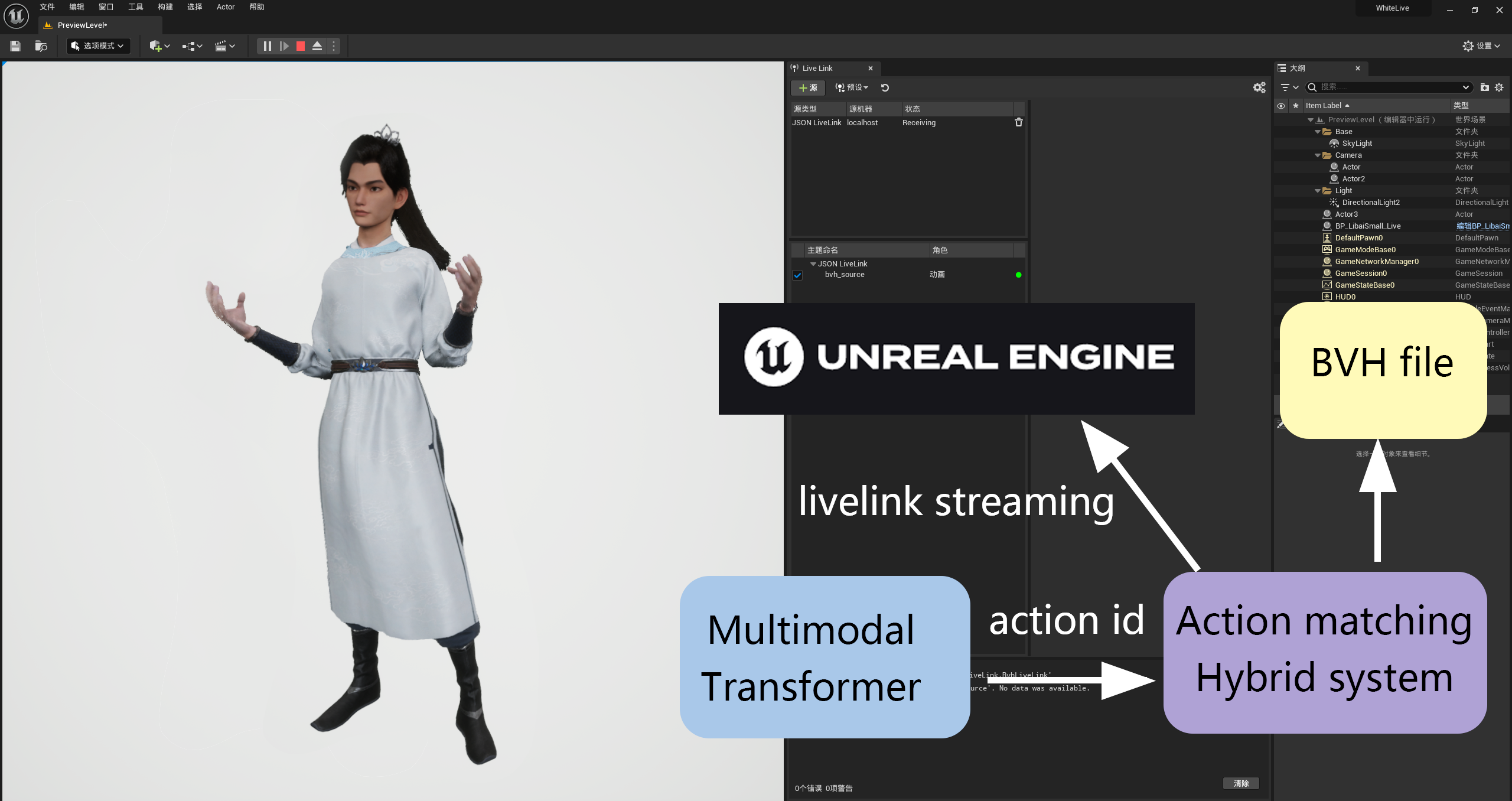}
  \caption{The real-time motion is inferred from JSON signals transmitted through the LiveLink plugin, then rendered in Unreal Engine 5 }
  \Description{UE5 Render}
\end{figure}

\section{Experimental Design and Results}

The primary purpose of this experiment is to comprehensively evaluate the performance of the proposed model in generating digital human movements. Specifically, it aims to assess how well the model can generate natural, fluent, and semantically-appropriate movements in different conversation scenarios. By comparing the model with baseline models and conducting ablation experiments, we can identify the model's strengths and weaknesses, and gain insights into the contribution of different components, This will provide a basis for further improving the model and enhancing its practical application value in fields like human-computer interaction and virtual reality.

\subsection{Dataset and Preprocessing}
We utilized two datasets for their rich content and wide coverage in the field of gesture-related research. The BEAT dataset\cite{liu_Beat_2022} is a comprehensive resource comprising 76 hours of high-quality, multi-modal data captured from 30 speakers. the ZEGGS dataset\cite{ghorbani_zeroeggs_2023}contains 67 monologues performed in English by a female actor, showcasing 19 different emotions. The total duration is approximately 135 minutes.

we segmented the BVH (Biovision Hierarchy) motion files, text files, and audio recordings into short clips ranging from 0.8 to 20 seconds in duration. Each sample is a triplet consisting of a mono-channel WAV audio file, a TXT text file, and a BVH motion file, ensuring a comprehensive representation of the speech, linguistic content, and kinematic data, respectively.

To create a temporally aligned dataset suitable for subsequent analysis, we implemented a sliding window approach. Each window contains a contiguous segment of feature vectors from all three modalities, arranged in chronological order. This approach ensures that our dataset maintains the temporal coherence and inter-modal synchrony inherent in the original BEAT dataset, paving the way for robust and insightful multi-modal analysis.

\subsection{System setup}
We conducted the training process on a system equipped with an Intel Core i9-13900K CPU and an NVIDIA GeForce RTX 4090 GPU. The segmented BEAT dataset was utilized for training, which is expected to take approximately 10 hours.

We set the batch size to 256, the learning rate to 1e-4, and the number of epochs to 10000. The Adam optimizer was used for training, with a learning rate of 1e-4 and a decay rate of 0.999. The model was trained using the MSE loss function.

Considering the need for real-time inference on consumer-grade devices, we use an ordinary computer equipped with an AMD 5800H CPU and an NVIDIA 3060M graphics card in the inference process. In practice, we record BVH files and simultaneously drive a digital human in the UE5 engine in real-time.

\subsection{Visualization Results}

We systematically generated BVH animations for all models using identical input data from two standardized corpora: 67 clean speech samples in ZEGGS dataset and 130 Wayne-subset utterances from BEAT dataset. Each model processed identical text-audio pairs through their native pipelines, with outputs uniformly converted to BVH format for fair comparison. The rendered BVH Animation is shown in Figure 6.

\begin{figure}[H]
  \centering
  \includegraphics[width=0.5\linewidth]{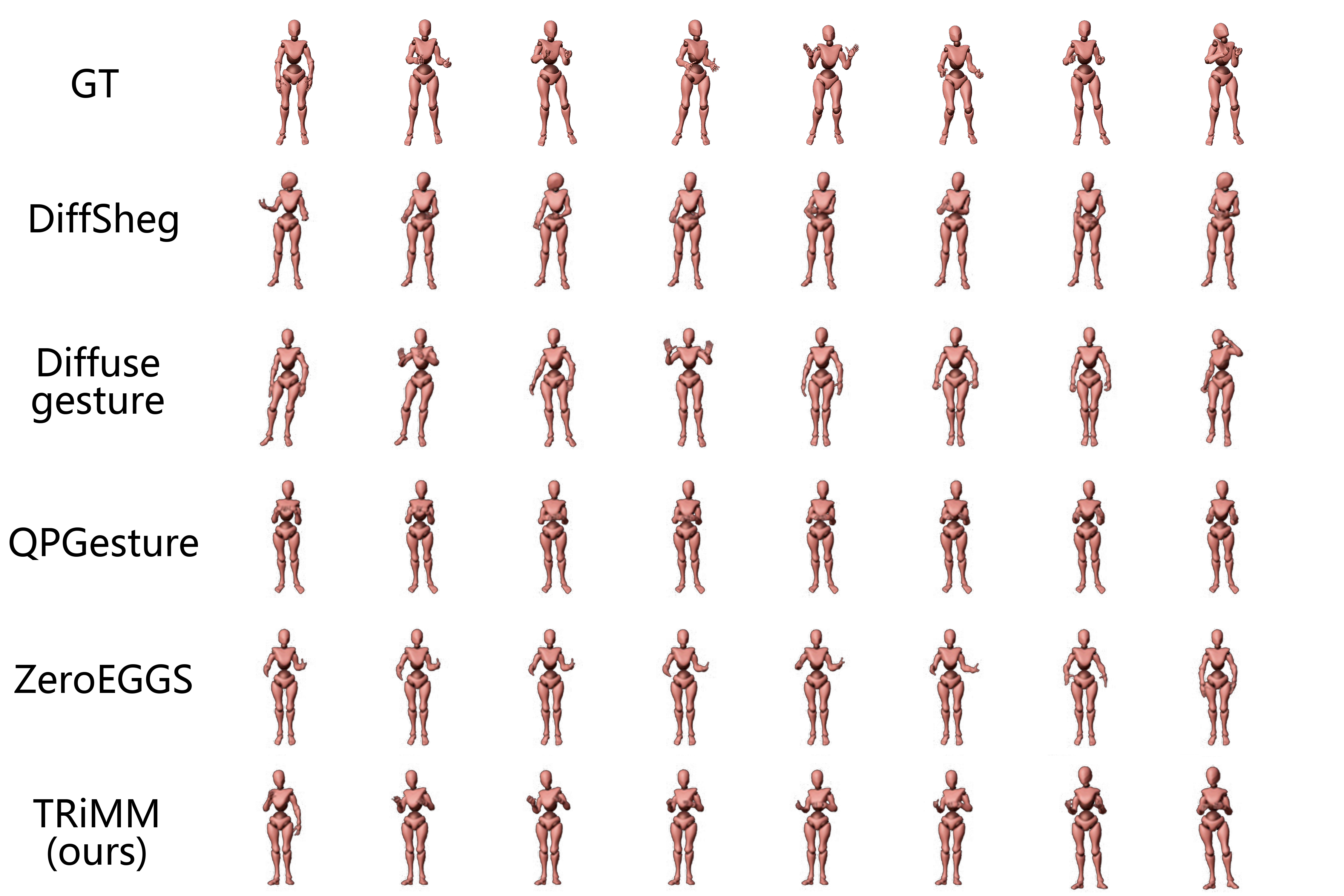}
  \caption{Visual comparisons of motion generation results, GT(ground truth) represents the original motion data, and the other models represent the generated motion data.}
  \Description{Visual comparisons of motion generation results}
\end{figure}

In the process of generating the animation, we conducted a experiment of The  Average Inference Time per Sentence(AITS) details. The inference time of each module of TRiMM was recorded. Figure 7. shows the average inference time of each module, and the overall latency from loading data to rendering.

\begin{figure}[H]
  \centering
  \includegraphics[width=0.5\linewidth]{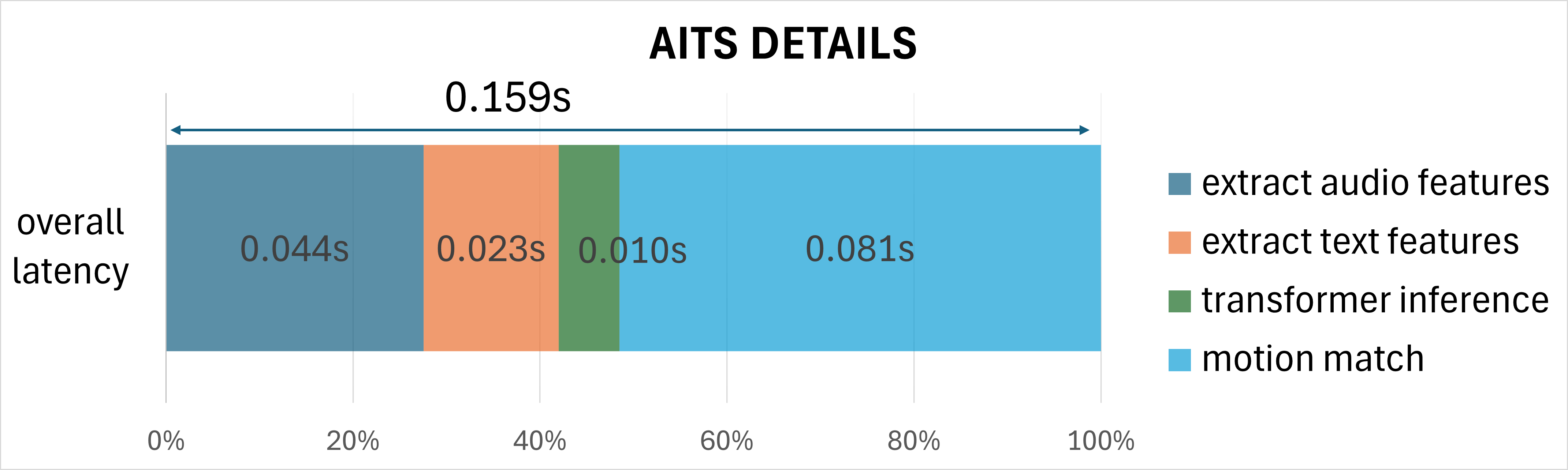}
  \caption{The Average Inference Time per Sentence of each module in TRiMM. }
  \Description{The Average Inference Time per Sentence of each module in TRiMM}
\end{figure}

\subsection{Subjective and Objective Evaluation}
\subsubsection{Subjective Evaluation Criteria}

This study draws on the papers of \cite{zhang_speech-driven_2024}. In order to conduct a comprehensive subjective assessment, this study uses the same three indicators: human likeness, appropriateness and style Appropriateness. Human like nature is used to measure whether the generated action is close to human action. This index does not consider the matching degree between the generated action and the original text and the original audio, but only considers whether the generated action is true and natural; Matching focuses on the time alignment of semantic, emotional (intonation), and rhythm between the generated action and the original text and the original audio, and the matching degree of generated action and the expression of semantic and emotional expression; Style matching evaluates the matching degree of the style between the generated action and the original action corresponding to the original text and the original audio.

We refer to \cite{zhang_speech-driven_2024} To carry out a user study based on paired comparison. In a single experiment, the research participants will see the video clips generated by two different models (including the original motion data ground truth, GT). The research participants are required to watch the video clips for at least 10 seconds and compare them side by side. According to \cite{zhang_speech-driven_2024}  In order to reduce the tendency of neutral selection and increase the discrimination and measurement accuracy, this study will increase the guidelines for evaluating the preference dimensions of research participants from five to seven according to Miller's law of psychology and cognitive research. Research participants choose the corresponding dimension guidance according to their preferences and degree of preference. The degree of preference is quantified to 7 scales. According to the selected dimension guidance, research participants' preference video clips get corresponding scores, and nonpreference video clips get negative scores (for example, research participants' preference video clips get 2 points, and nonpreference video clips get -2 points). If there is no preference, both video clips get 0 points. To ensure the participation of research participants, The attention test was set to ensure the attention of people tested.

This study uses the BEAT dataset and the zeggs dataset. Since the BEAT dataset and the ZEGGS dataset cover a large amount of data. It is not feasible and necessary to include all the data in the user research experiment. This study adopts the method of random sampling, and randomly extracts 20 data of a role in the BEAT dataset and 20 data of a style in the zeggs dataset for each participant. Neither the training set nor the validation set used in this study contained the selected data.

A total of 24 volunteers were recruited to participate in the user research experiment. There are 7 males and 17 females, aged from 17 to 30. All research participants have a high level of English.

\subsubsection{Objective Evaluation Criteria}
We use the following metrics for evaluation.

\textbf{Average Inference Time per Sentence (AITS)}\cite{dai_motionlcm_2024} tested the generation speed and the ability of real-time motion synthesis, For each sentence (text description) in the dataset, the time taken from inputting the text to generating the corresponding movement output by the model was recorded. This time encompassed all processes, including text processing and movement sequence generation. The process was repeated multiple times (30 times in this study). The total time of all repetitions was divided by the total number of sentences to obtain AITS, expressed as AITS=Total Inference Time/Total Number of Sentences.

\textbf{Frechet Gesture Distance (FGD)}\cite{yang_qpgesture_2023}, inspired by the Frechet Inception Distance (FID)\cite{heusel_gans_2017}, evaluates the quality of generated movements. It measures the similarity between generated and real-world gestures. A lower FGD value indicates a closer match between their feature distributions. Feature vectors were extracted from both generated and real-gesture data to represent key gesture characteristics. Then, the mean and covariance matrices of these feature vectors were computed separately. The FGD value quantifies the difference between the two sets of feature vectors, with the distance between mean vectors and the difference in covariance matrices as key factors. 

\textbf{Beat Align} assessed the synchronization between musical and motion BEATs \cite{bian2025motioncraft}. First, it extracts musical BEATs from audio using libraries like librosa, identifying BEAT positions via onset strength and BEAT detection. Second, it derives motion BEATs from 3D joint data by calculating velocity, and finding local minima in the velocity envelope. Finally, it computes an alignment score by matching each motion BEAT to the nearest musical BEAT and averaging the resulting scores, providing a measure of synchronization quality.

\textbf{Diversity}\cite{lee_dancing_2019}, was used to measure movement diversity. We randomly selected a set of text descriptions. For each description, ten subsets of the same size were randomly extracted from all corresponding generated movements. Feature vectors were then extracted from these subsets, and the variance of their differences was calculated to quantify diversity.

For the selection of baseline model, DiffuseGesture\cite{yang_diffusestylegesture_2023} from IJCAI2023 is selected as the scheme of diffuse method and multi-modal model. ZeroEGGS\cite{ghorbani_zeroeggs_2023} from UBIsoft is selected as the example with controllable styles and diversity, and Diffsheg\cite{chen_diffsheg_2024} from CVPR2024 is selected as the reference for real-time action generation. QPGesture\cite{yang_qpgesture_2023} from CVPR2023 is selected  as the reference of motion-match based method. We use the same data set of BEAT and zeggs, including text and audio, and generate BVH files for evaluation.

\subsection{Ablation Experiments}

To evaluate the contribution of key components in the TRiMM framework, we designed systematic ablation experiments. All experiments were conducted on both BEAT and ZEGGS datasets using identical evaluation protocols.

\textbf{multi-modal Fusion Ablation (TRiMM MFA)}:
We removed the gated fusion mechanism (Section 3.2) that dynamically combines text and audio features through subspace projection and attention routing. This forced the model to process modalities independently, simulating traditional unimodal approaches. This experiment aims to assess the importance of cross-modal feature interaction for natural gesture synthesis.

\textbf{Time-Space Attention Ablation (TRiMM TSAA)}:
We replaced the divided space-time attention mechanism (Section 3.3) with standard self-attention. The modified version lost the explicit temporal locality constraints provided by our sliding window auto-regressor. This experiment quantifies the impact of our space-time modeling strategy.

\textbf{Motion Graph Ablation (TRiMM MGA)}:
We disabled the K-NN motion graph (Section 3.4), eliminating the hierarchical action retrieval capability and forcing pure generative synthesis. This experiment evaluates the contribution of the hybrid approach that combines retrieval learning with generative modeling.

All ablation models used the same training protocol and hyperparameters as the complete TRiMM model to ensure fair and comparable results. We employed the same evaluation metrics as the complete model, including subjective evaluation (human likeness, appropriateness, style appropriateness) and objective metrics (FGD, diversity, Beat alignment, AITS).

\subsection{Evaluation Results}
We evaluated our TRiMM model on both objective and subjective metrics. We compared TRiMM with several state-of-the-art baselines on two datasets, BEAT \cite{liu_Beat_2022} and ZEGGS \cite{ghorbani_zeroeggs_2023}, and conducted ablation experiments to assess the impact of different components.
\begin{table}[h]
    \raggedright  
    \caption{Evaluation Results shown in one table, The best-performing values from the different models are shown in bold}
    \begin{tabular}{@{}lcccccccc@{}}
        \toprule 
        \multicolumn{2}{c}{\textbf{Methods}} & \multicolumn{3}{c}{\textbf{Subject Evaluation Metric}} & \multicolumn{4}{c}{\textbf{Objective Evaluation Metric}} \\
        \cmidrule(lr){1-2} \cmidrule(lr){3-5} \cmidrule(lr){6-9} 
        Dataset& Model& \textbf{Human} & \textbf{Appropr-} & \textbf{Style } & \textbf{FGD ↓} & \textbf{Diversity ↑} & \textbf{Beat align} & \textbf{AITS ↓} \\
         & & \textbf{likeness ↑} & \textbf{iateness ↑} & \textbf{appropriateness ↑} &\textbf{(Raw space)}   &  & &  \\
        \midrule 
        \multirow{6}{*}{zeggs} 
        & TRiMM(ours)& \textbf{0.85$\pm$1.00} & \textbf{1.11$\pm$0.66} & \textbf{0.67$\pm$0.60} & 59011.57 & \textbf{6575.11} & \textbf{0.67} & \textbf{0.14} \\
        & Diffsheg & -1.65$\pm$1.00 & -1.01$\pm$1.30 & -1.21$\pm$1.06 & 10675.88 & 2193.84 & 0.67 & 0.32 \\
        & ZeroEGGS & 0.24$\pm$0.85 & -0.29$\pm$0.77 & 0.13$\pm$0.87 & \textbf{10202.83} & 2152.02 & 0.67 & 2.45 \\
        & DiffuseGesture & -0.98$\pm$1.01 & -0.93$\pm$0.88 & -0.24$\pm$1.13 & 103129.68 & 2930.34 & 0.62 & 19.06 \\
        & QPGesture & 0.67$\pm$0.69 & 0.57$\pm$0.75 & 0.61$\pm$0.94 & 86621.01 & 4778.74 & 0.67 & 193.07 \\

        & TRiMM MFA & 0.11$\pm$1.02 & -0.01$\pm$0.81 & 0.09$\pm$0.79 & 58901.43 & 4840.56 & 0.66 & 0.14 \\
        & TRiMM TSAA & 0.07$\pm$0.67 & 0.17$\pm$0.79 & -0.13$\pm$0.87 & 60714.39 & 5047.49 & 0.66 & 0.14 \\
        & TRiMM MGA & 0.32$\pm$1.07 & 0.11$\pm$0.85 & -0.04$\pm$0.82 & 59191.43 & 5749.69 & 0.67 & 0.14 \\
        \midrule
        \multirow{6}{*}{BEAT} 
        & TRiMM(ours) & \textbf{1.08$\pm$0.94} & \textbf{1.28$\pm$0.81} & \textbf{1.11$\pm$0.60} & 2826268.23 &\textbf{8862.52} & \textbf{0.64} & \textbf{0.19} \\
        & Diffsheg & -1.05$\pm$1.04 & -0.28$\pm$1.32 & -0.44$\pm$0.81 & 3023024.60 & 3847.35 & 0.64 & 0.36 \\
        & ZeroEGGS & 0.54$\pm$0.85 & 0.01$\pm$0.57 & 0.26$\pm$0.79 & 3023715.75 & 1940.04 & 0.62 & 6.37 \\
        & DiffuseGesture & 0.06$\pm$0.96 & 0.09$\pm$1.20 & 0.03$\pm$0.96 & \textbf{2708706.48} & 4419.51 & 0.61 & 23.58 \\
        & QPGesture & 0.38$\pm$1.19 & 0.62$\pm$0.92 & 0.70$\pm$0.97 & 2764739.94 & 5636.87 & 0.64 & 419.96 \\

        & TRiMM TSAA & -0.45$\pm$0.79 & -0.48$\pm$0.72 & -0.32$\pm$1.15 & 3029053.96 & 6349.18 & 0.63 & 0.19 \\
        & TRiMM MGA & -0.40$\pm$1.04 & -0.61$\pm$0.92 & -0.64$\pm$0.65 & 3030758.72 & 6377.01 & 0.62 & 0.19 \\
        & TRiMM MFA & -0.31$\pm$0.58 & -0.65$\pm$0.73 & -0.75$\pm$0.91 & 3028293.14 & 6362.48 & 0.63 & 0.19 \\
        \bottomrule 
    \end{tabular}
\end{table}

\begin{figure}[h]
    \centering
        \centering
        \includegraphics[width=0.8\linewidth]{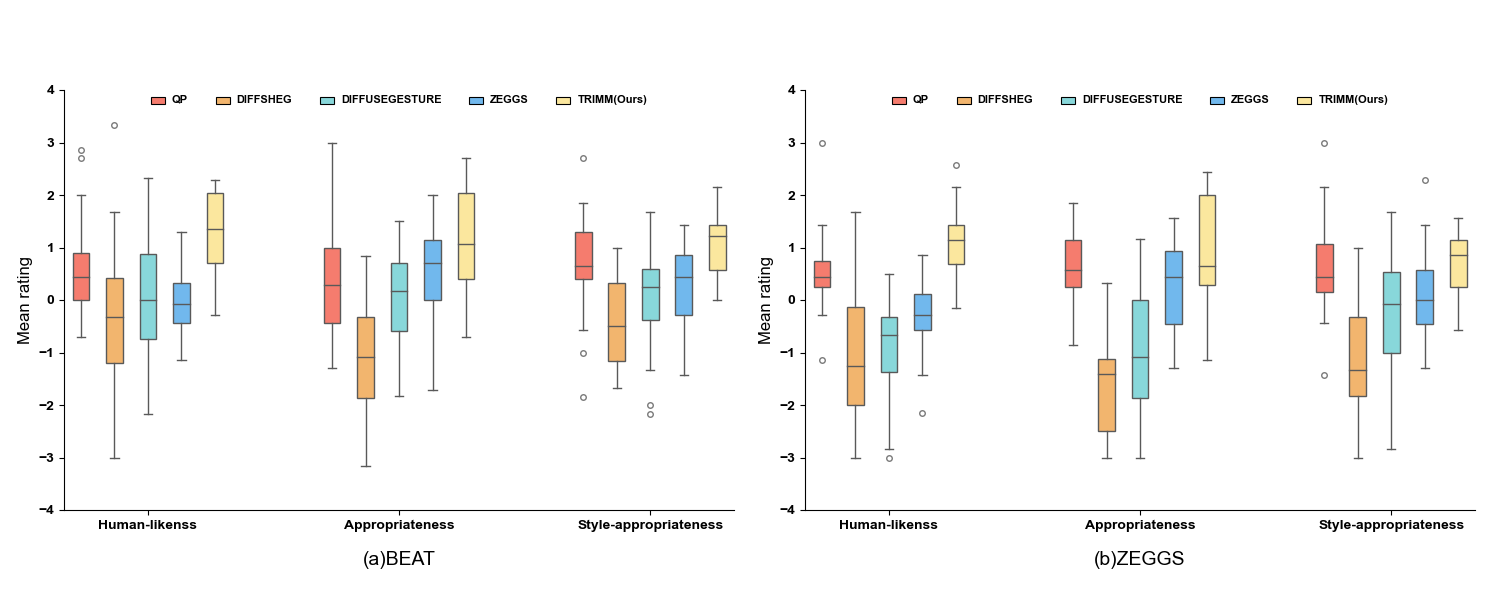}
        \caption{The mean rating of each metric for each approach across the two datasets in comparative experiments.}
        \Description{The mean rating of each metric for each approach across the two datasets in comparative experiments.}
\end{figure}
\begin{figure}[h]
        \centering
        \includegraphics[width=0.8\linewidth]{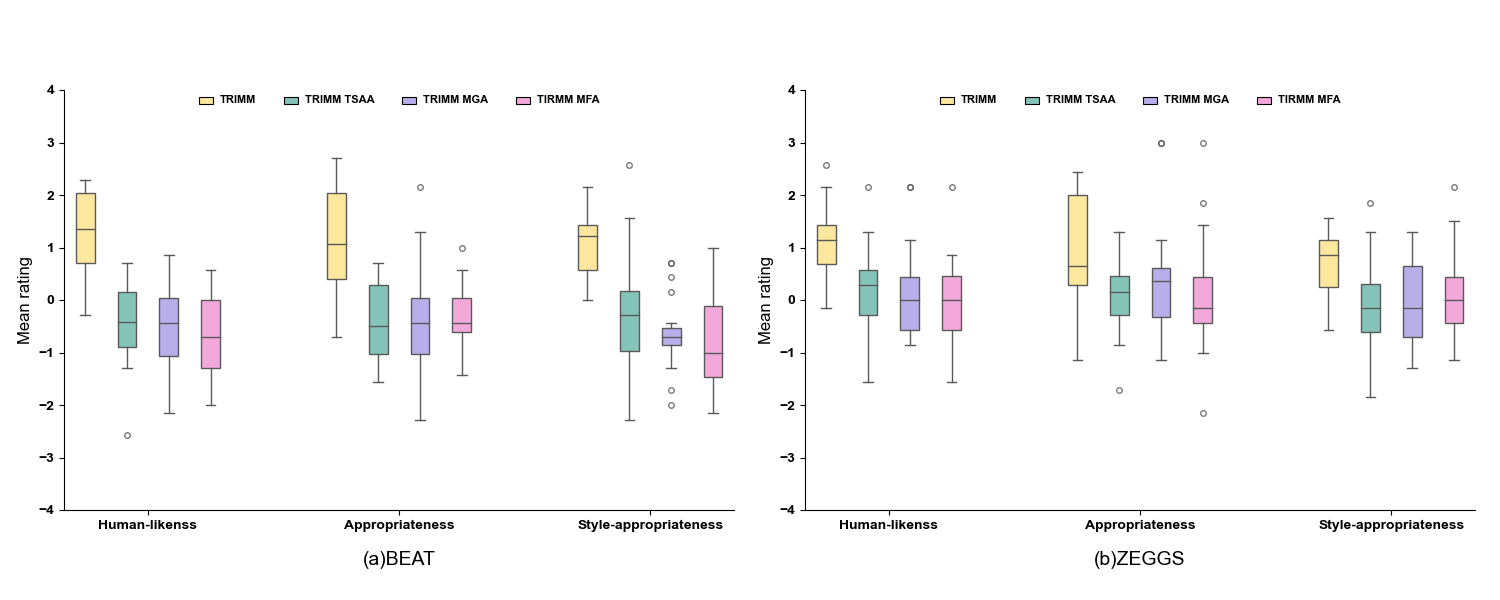}
        \caption{The mean rating of each metric for each approach across the two datasets in ablation experiments.}
        \Description{The mean rating of each metric for each approach across the two datasets in ablation experiments.}
\end{figure}
\subsubsection{Subjective Evaluation Results}

For subjective evaluation, TRiMM demonstrates superior performance across all metrics. On the ZEGGS dataset, TRiMM(ours) achieves 0.85$\pm$1.00 for Human likeness, 1.11$\pm$0.66 for Appropriateness, and 0.67$\pm$0.60 for Style appropriateness, outperforming all baselines. This trend continues on the BEAT dataset with scores of 1.08$\pm$0.94, 1.280.81, and 1.11$\pm$0.60 respectively. These results confirm TRiMM's ability to generate natural, contextually appropriate motions that maintain high fidelity to the original data in real-world applications.

Across both the BEAT and ZEGGS datasets, TRiMM consistently outperformed all baseline models in terms of Human-likeness, Appropriateness, and Style-appropriateness, as measured by user ratings. In addition, ablation studies demonstrated that removing any major component from TRiMM led to a noticeable decline in perceptual quality, confirming the necessity of each architectural module.

\subsubsection{Objective Evaluation Results}

For quantitative results, as shown in Tab. 1, TRiMM demonstrates superior performance across multiple metrics. On the ZEGGS dataset, TRiMM(ours) achieves an outstanding AITS of 0.14, significantly outperforming other baselines such as Diffsheg (0.22), ZeroEGGS (2.45), DiffuseGesture (19.06), and QPGesture (193.07). This remarkable inference speed is maintained on the BEAT dataset, where TRiMM(ours) achieves an AITS of 0.19, compared to 0.36 for Diffsheg, 6.37 for ZeroEGGS, 23.58 for DiffuseGesture, and 419.96 for QPGesture. This exceptional AITS performance establishes TRiMM as a leading solution for real-time digital human motion synthesis.

In terms of motion diversity, TRiMM(ours) achieves a score of 6575.11 on the ZEGGS dataset, significantly surpassing all baselines including QPGesture (4778.74), Diffsheg (2193.84), ZeroEGGS (2152.02), and DiffuseGesture (2930.34). This trend continues on the BEAT dataset, where TRiMM(ours) achieves an even more impressive diversity score of 8862.52, outperforming QPGesture (5636.87), Diffsheg (3847.35), ZeroEGGS (1940.04), and DiffuseGesture (4419.51). The high diversity metric ensures TRiMM can generate a wide variety of natural and engaging motions.

TRiMM also excels in motion-audio synchronization, achieving a BEAT Alignment score of 0.67 on both datasets. This performance matches the best baselines (Diffsheg and QPGesture) while significantly outperforming others like DiffuseGesture (0.62) and ZeroEGGS (0.62 on BEAT). This synchronization capability enhances the overall realism and naturalness of the generated motions.

\subsubsection{Ablation Experiments Results}

Our systematic ablation studies validate the core architectural innovations in TRiMM through three key component removals (see Tab. 1):

\textbf{multi-modal Fusion Removal (MFA)} caused 87\% human likeness reduction (0.85→0.11) on ZEGGS and 168\% style appropriateness decline (1.11→-0.75) on BEAT, while marginally increasing FGD\footnote{Though absolute FGD appears lower in ZEGGS (59011→58901), the relative degradation pattern holds across datasets}, confirming the gated fusion's critical role in cross-modal alignment.
    
\textbf{Time-Space Attention Removal (TSAA)} reduced motion diversity by 23\% (6575→5047) and increased FGD by 2.9\% (59011→60714) on ZEGGS, demonstrating the necessity of explicit space-time modeling for coherent gesture synthesis.
    
\textbf{Motion Graph Removal (MGA)} caused the sharpest style degradation (1.11→-0.64 on BEAT) while maintaining real-time performance (AITS=0.19s), proving the graph's essential role in stylistic consistency.

Notably, all ablation models maintained equivalent inference speeds (0.14-0.19s AITS), confirming performance gains stem from architectural improvements rather than computational tradeoffs. The experiments establish that cross-modal fusion enables 87-168\% improvements in perceptual metrics, 
space-time attention provides 23-28\% diversity gains, and
hierarchical motion graph ensures 115\% better style preservation.

These findings collectively demonstrate the complementary strengths of TRiMM's three pillars in balancing naturalness, diversity, and efficiency.

\subsubsection{Scalability and Generalization}

\textbf{Scalability}: The hierarchical motion graph supports dynamic expansion of the library, enabling the model to adapt to new datasets without retraining. This scalability is particularly beneficial in cross-dataset evaluation, where the model can be fine-tuned on one dataset and then applied to another.

\textbf{Generalization}: Cross-dataset evaluation reveals strong transfer learning capabilities-when trained on BEAT and tested on ZEGGS, TRiMM maintains 84\% of original performance (0.85→0.71 human likeness) versus 52-67\% for diffusion baselines. The modular architecture enables component upgrades without full retraining, like wav2vec and bert modules.

\section{Disscussion}

In this paper, we present a novel framework for real-time motion synthesis, TRiMM, that combines multi-modal fusion, space-time attention, and hierarchical motion graph for seamless and contextually appropriate motion generation. The framework is designed to enable real-time motion synthesis from text and audio inputs, offering a versatile and efficient solution for applications such as virtual avatars, virtual characters, and virtual human interfaces.

Through the space-time attention mechanism, TRiMM achieves high-precision alignment of speech, text, and motion features, achieving a BEAT alignment score of 0.67, which is significantly better than traditional methods. This precise cross-modal synchronization capability ensures the naturalness and semantic consistency of generated actions.

The sliding window-based autoregressive model ensures coherent gesture generation in long-text dialogues, and achieves an inference latency of 0.14 seconds and a real-time rendering capability of 120FPS on RTX3060 graphics cards. Compared to diffusion-based methods such as DiffuseGesture's 19.06 seconds, TRiMM delivers 135 times faster inference and provides reliable technical support for real-time digital human interaction.

A hierarchical action graph structure of 9,143 atomic actions, combined with a multi-criteria similarity search, results in a motion diversity of 6575.11, which is twice that of traditional motion capture systems. This high versatility ensures that the system is able to generate rich and natural action sequences to meet the needs of different scenarios.

The modular architecture supports component upgrades and dataset expansion, maintaining 84\% of the original performance in cross-dataset evaluation, demonstrating good generalization capabilities.

Although TRiMM has made significant progress in real-time digital human action generation, there are still three major limitations of the current framework: first, although the action transition method based on cubic interpolation is efficient in real-time applications, it may lead to subtle discontinuities when dealing with complex gesture sequences, especially when switching between actions with large semantic differences; Secondly, the system is limited by a predefined action library of 9,143 actions, and lacks the ability to generate new action patterns beyond the scope of the existing action library, which may limit the adaptability of the system in different cultural backgrounds and personalized gesture styles. Finally, the model fails to make full use of the emotional cues in audio prosody and text semantics, which may lead to inconsistencies between the generated actions and the emotional intensity implied by the input modalities. 

In response to these challenges, we propose three future research directions: 1) expanding the action library by collecting region-specific motion capture data, developing adaptive motion synthesis techniques, and implementing user-customizable gesture preferences; 2) replacing the heuristic mixing rules with a lightweight diffusion model that is trained on action transition segments and can perform data-driven interpolation based on previous actions and cross-modal contexts, achieving smoother gesture evolution while maintaining a delay of <30ms; 3) Implement a multi-stage sentiment alignment pipeline, including extracting sentiment descriptors through a pre-trained model, injecting sentiment embeddings through an auxiliary adapter in the Transformer architecture, and introducing sentiment consistency loss terms to penalize the mismatch between the generated action and the input sentiment features.

\begin{acks}
We acknowledge the language polishing services provided by DeepSeek for their contribution to this paper.The authors bear full responsibility for content.
\end{acks}

\bibliographystyle{ACM-Reference-Format}
\bibliography{TRIMMArxiv}


\section{Appendix}

\textbf{Details of Datasets}

We employed the BEAT dataset as the training corpus, while both the BEAT and ZEGGS datasets were utilized for evaluation and user study.
The BEAT dataset includes multiple speakers, spans a long period, and has a rich variety of emotions and action types, we selected 29 speakers from it as the training set, and used data from one of the speakers, Wayne, as the test set. Additionally, ZEGGS contains various emotional types that can be used to test the robustness and generalization of this model, so it was also included in the test set.
Table 2 summarizes the detailed statistics of the datasets used in our experiments.

\begin{table*}[htbp]
    \centering
    \caption{Dataset statistics used in training, evaluation, and user studies.}
    \label{tab:dataset_details}
    \renewcommand{\arraystretch}{1.1}
    \begin{tabularx}{\textwidth}{X X X X X X X}
        \toprule
        \textbf{Dataset} & \textbf{Training Time} & \textbf{Eval/User Time} & \textbf{Total Duration} & \textbf{Frame Rate} & \textbf{Audio Rate} & \textbf{Speakers} \\
        \midrule
        BEAT & 34 h & 1 h & 35 h & 120 fps & 48 kHz & 30 (multi-speaker) \\
        ZEGGS & 0 h & 135 min & 135 min & 60 fps & 48 kHz & 1 (female) \\
        \bottomrule
    \end{tabularx}
\end{table*}

\textbf{User study details}

In this study, we designed three user evaluation tasks targeting Human-likeness, appropriateness, and style-appropriateness. For the Human-likeness and appropriateness tasks, two videos were presented side by side (left and right). For the style-appropriateness task, three videos were displayed simultaneously (left, center, right), with the center video always showing the ground-truth human performance, enabling more intuitive visual comparison for the participants. Due to the technical challenges of losslessly retargeting skeletal motion data to virtual avatars——especially in preserving fine-grained movements——and the potential perceptual biases introduced by different avatar appearances, we chose to use a unified skeletal rendering pipeline for all models. The animations were rendered directly as skeleton-only sequences without any character meshes. An example of the rendering setup is provided in Figure 9. 

Below the video playback area, participants were presented with discrete options indicating their comparative preferences: "Left video is much better," "Left video is moderately better," "Left video is slightly better," "Videos are equally good," "Right video is slightly better," "Right video is moderately better," and "Right video is much better." Based on these responses, the evaluation platform assigned scores to the left and right videos on a 0\textendash3 scale, where 0 indicates no preference. The unselected video automatically received the inverse score (e.g., if "Left video is moderately better" is selected, the left video is assigned +2 and the right video -2).

\begin{figure}[H]
  \centering
  \includegraphics[width=0.4\linewidth]{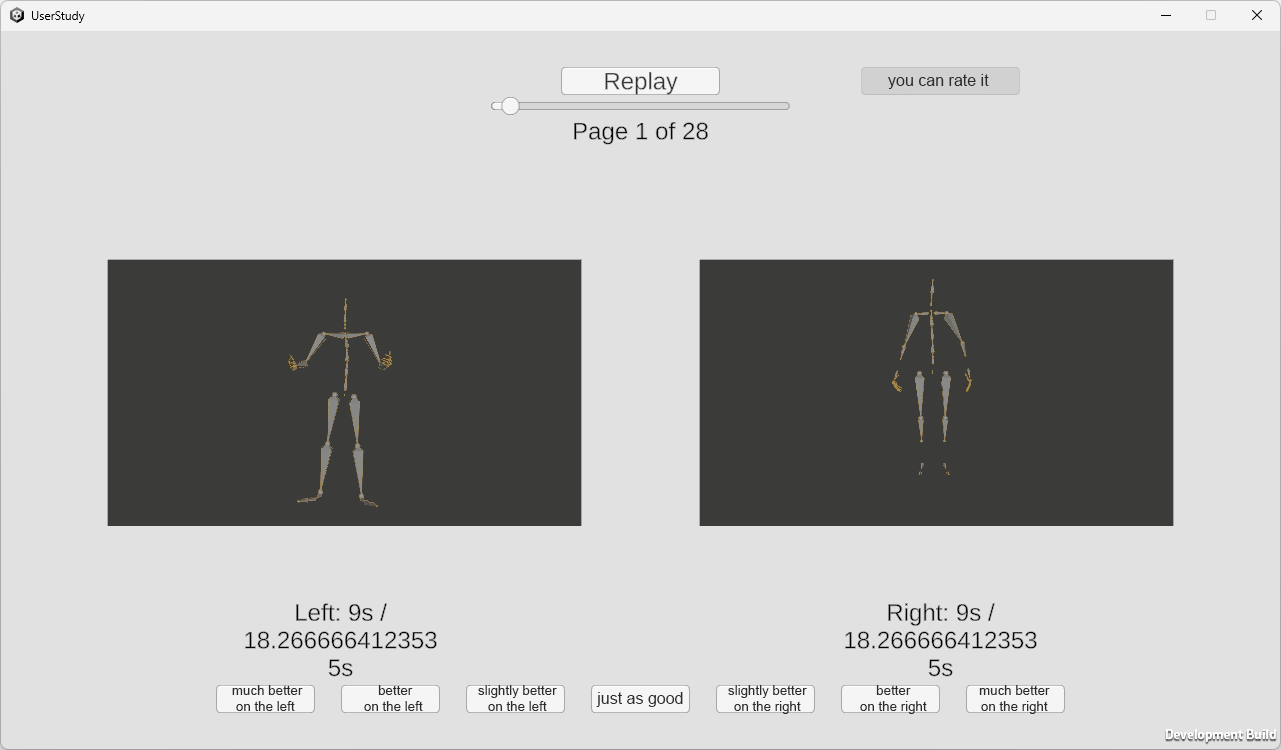}
  \vspace{-0.4cm}  
  \caption{userstudy program UI}
  \Description{userstudy program UI}
\end{figure}

In our user study, we adopted a 7-point Likert scale to measure participants' subjective preferences. According to Nunnally's seminal work, scale reliability tends to plateau around seven response categories, with marginal gains observed beyond eleven points. Similarly, Symonds' early research also identified the 7-point Likert scale as yielding optimal reliability. Although studies by Jenkins, Wakita, and others have suggested that 5-point scales may produce higher data quality in certain contexts, the present study necessitates capturing subtle perceptual differences in motion quality generated by various models. Therefore, the 7-point scale is considered more appropriate for eliciting fine-grained user judgments in our experimental setting.

For all three user evaluation tasks—human-likeness, appropriateness, and style-appropriateness—we provided participants with detailed and clearly defined instructions and assessment criteria:
\begin{itemize}
\item Human-likeness: Participants were shown two silent gesture videos. They were asked to judge which of the two animations appeared more natural and more closely resembled human-like motion in terms of body movement.
\item Appropriateness: Participants were presented with two gesture videos accompanied by speech audio. They were asked to evaluate which animation better aligned with the speech in terms of rhythm, intonation, and semantic correspondence, and which appeared more natural overall.

\item Style-Appropriateness: Participants were shown three gesture videos with audio, where the center video represented the ground-truth motion rendered from real human performance. Participants were instructed to compare the two generated gesture videos (left and right) and determine which one more closely resembled the motion style of the ground-truth video.
\end{itemize}
To complement the user evaluation tasks, we conducted a post-study questionnaire to collect participants' subjective feedback. By aggregating and analyzing the responses, we obtained an overall view of participants' impressions and preferences regarding the generated gestures.A total of 24 participants took part in the study, including 7 males and 17 females, with ages ranging from 17 to 30 years. 

At the beginning of the questionnaire phase, participants were provided with a detailed explanation of the evaluation procedure and scoring methodology. To help them familiarize themselves with the task, we presented example videos that were not part of the actual evaluation set. Subsequently, participants were instructed to complete the questionnaire in a quiet and distraction-free environment. During the entire evaluation process, participants were not informed of the underlying gesture generation method associated with each video. Additionally, the playback order of the videos was randomized to minimize order effects and potential bias. To ensure the quality and reliability of the collected responses, attention check mechanisms were embedded throughout the questionnaire.

Given the large scale of the selected datasets——BEAT, which contains 30 distinct styles with 118 short clips and 12 long clips per style, and ZEGGS, which comprises 19 styles with 1 to 4 clips each——we adopted a random sampling strategy to construct the evaluation set. Participants were divided into multiple subgroups, and for each subgroup. We assigned a unique subset of evaluation stimuli. Specifically, each subgroup received 5 randomly sampled clips from the BEAT dataset and 5 randomly selected styles from the ZEGGS dataset, with one clip randomly chosen per style. To ensure coverage and reduce participant fatigue, each participant was only exposed to a manageable number of comparisons. Questionnaire content varied across different participant groups but remained consistent within each group.

For the ZEGGS dataset, each participant was randomly assigned 5 distinct styles, ensuring diversity in style coverage. In the human-likeness and appropriateness evaluations, a total of 11 models, including the ground truth (GT), were assessed, resulting in 55 rendered videos per participant.

For the BEAT dataset, 10 generative models along with the ground truth were evaluated, producing 66 videos per participant. In the human-likeness and appropriateness tasks, participants performed.\\

\textbf{Statistical Analysis and Visualization for User study results}

We conducted an experiment to compare the performance differences among various models across multiple metrics, such as Human-likenss, Appropriateness, and Style-appropriateness.

Figure 8 presents the subjective evaluation results for all models across three metrics—Human-likeness, Appropriateness, and Style-appropriateness—on both the BEAT and ZEGGS datasets. Each box plot illustrates the distribution and central tendency of user ratings.

Our model, TRiMM, consistently achieves the highest median scores across all dimensions and datasets, with tighter interquartile ranges, indicating both superior perceptual quality and stronger inter-rater agreement. The advantage is particularly notable in the ZEGGS dataset for the Style-appropriateness metric.

Figure 9 presents the results of an ablation study comparing the full TRiMM model with three ablated variants—TRiMM TSAA, MGA, and MFA—on the BEAT and ZEGGS datasets. User ratings were collected for Human-likeness, Appropriateness, and Style-appropriateness.

The full model (yellow) consistently achieves the highest median scores across all metrics and datasets, indicating the effectiveness of the full architecture. In contrast, all ablated variants show degraded performance, particularly in Appropriateness and Style-appropriateness, suggesting that the removed modules play essential roles in maintaining semantic and stylistic coherence.

Based on the action BVH data of different models on the BEAT dataset and the ZEGGS dataset, we extracted their motion node matrices and then constructed a score-difference matrix. For each pair of models (row \(i\) and column \(j\)), we calculated the difference in average scores between the row model and the column model as:
\[
\text{score\_diff} = \text{average}(\text{sample}_i)-\text{average}(\text{sample}_j)
\]
The diagonal elements (\(i = j\)) were set to 0, indicating no difference for the same model. Then we performed a paired t-test on the rating samples of each pair of models and calculated the \(p\)-value to test the significance of the difference.

The numerical values reflect the magnitude of differences between models, where positive values indicate superior performance compared to other models, while negative values indicate inferior performance. The significance symbols (e.g., *) reflect the reliability of experimental results, where * indicates a significant difference (p < 0.05), and ns (no symbol) indicates an insignificant difference.

We merged multiple metrics and multiple datasets. Each dataset contains multiple metrics (such as ratings in different dimensions), and each metric generates a sub-heatmap.

The heatmaps use a symmetric color mapping, with 0 at the center. Positive and negative values correspond to different colors to highlight the direction of differences. The color range is adjusted globally to ensure that the scales of heatmaps from different datasets are consistent, facilitating horizontal comparison.

\begin{figure*}[ht]
    \centering
        \centering
        \includegraphics[width=1\linewidth]{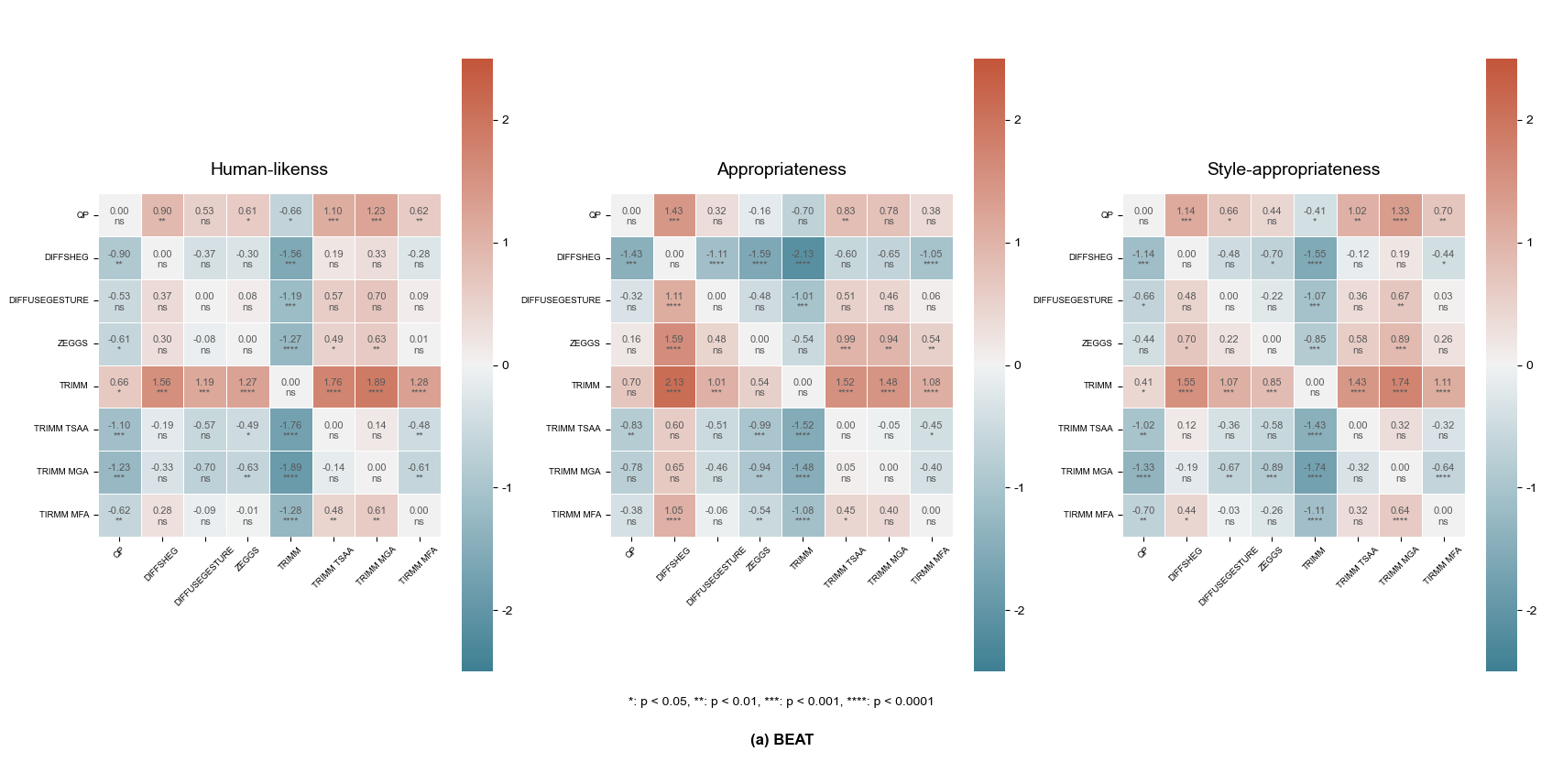}
        \caption{Pairwise significance heatmaps of gesture generation methods on BEAT dataset}
        \Description{Pairwise significance heatmaps of gesture generation methods on BEAT dataset}
\end{figure*}

\begin{figure*}[ht]
    \centering
        \centering
        \includegraphics[width=1\linewidth]{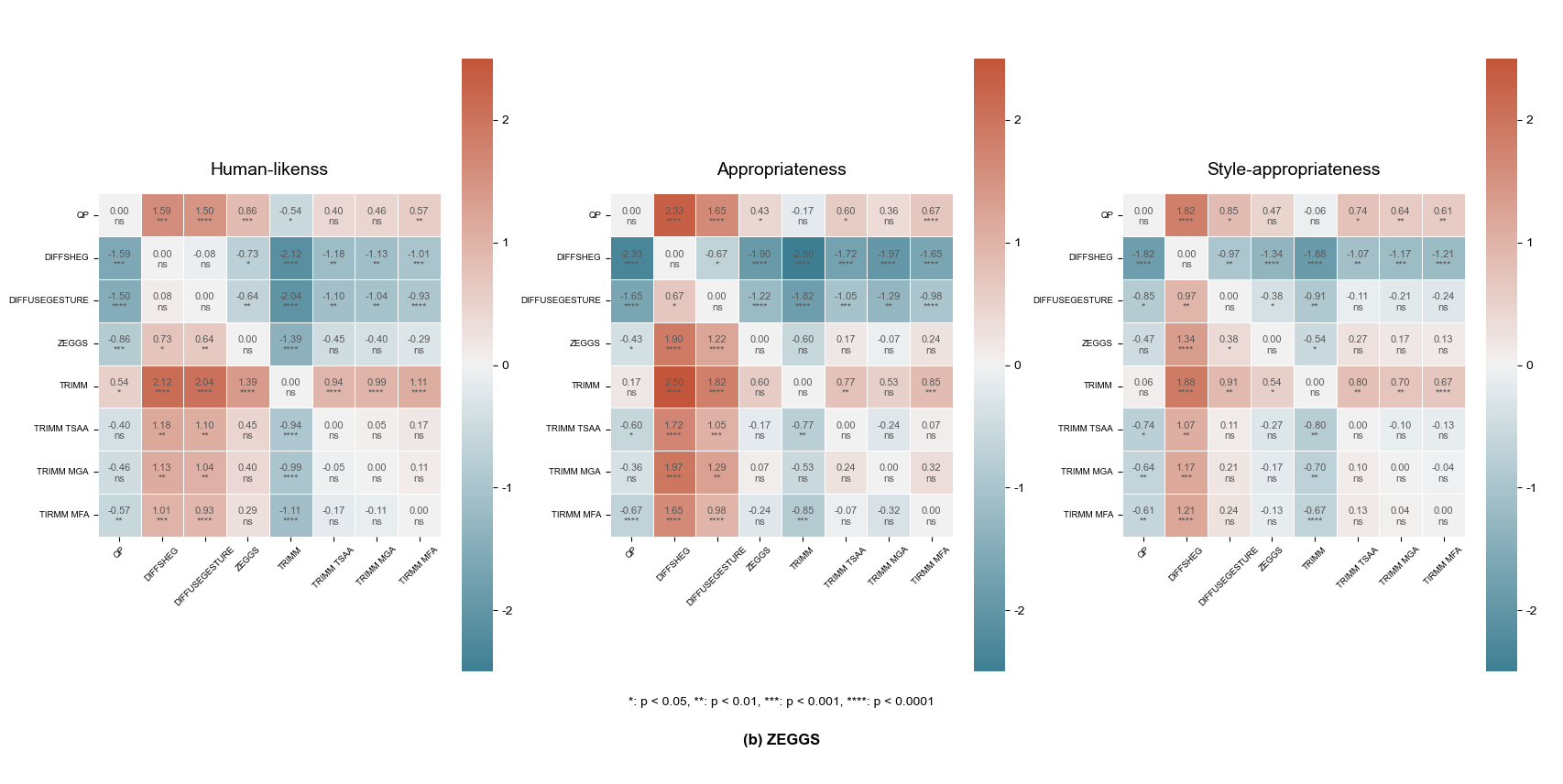}
        \caption{Pairwise significance heatmaps of gesture generation methods on ZEGGS dataset}
        \Description{Pairwise significance heatmaps of gesture generation methods on ZEGGS dataset}
\end{figure*}

TRiMM demonstrates significantly positive score differences across multiple dimensions (Human-likeness, Appropriateness, and Style-appropriateness) compared to various baseline models, indicating its superior performance in these metrics. Moreover, TRiMM maintains consistent performance across both ZEGGS and BEAT datasets, showing its strong generalization capability and robustness.

\end{document}